\documentclass[%
reprint,
amsmath,amssymb,
aps,
pra,
superscriptaddress,nofootinbib,twocolumn
]{revtex4-2}

\usepackage[USenglish]{babel}
\usepackage[T1]{fontenc}
\usepackage{hyperref}
\usepackage{graphicx}
\usepackage{dcolumn}
\usepackage{bm}
\usepackage{braket}
\usepackage{xcolor}
\usepackage{bbold}

\usepackage{xspace}
\usepackage{siunitx}
\usepackage{xfrac}
\usepackage{hyperref}
\usepackage[nameinlink]{cleveref}
\usepackage{appendix}

\usepackage{xifthen}
\usepackage{xcolor}
\hypersetup{
	colorlinks,
	linkcolor={red!75!black},
	citecolor={blue!75!black},
	urlcolor={blue!75!black}
}

\begin{document}
	
	\preprint{APS/123-QED}
	
    \title{Harmonically trapped fermions in one dimension: \\ A finite temperature lattice Monte Carlo study}% Force line breaks with \\
	
	\author{Felipe Attanasio}
	\author{Marc Bauer}
	\email{bauer\_m@thphys.uni-heidelberg.de}
    \author{Renzo Kapust}
	\affiliation{
		Institute for Theoretical Physics, Universit\"{a}t Heidelberg, Philosophenweg 16, D-69120, Germany
	}

	\author{Jan M. Pawlowski}
	\affiliation{
		Institute for Theoretical Physics, Universit\"{a}t Heidelberg, Philosophenweg 16, D-69120, Germany
	}%
	\affiliation{
		ExtreMe Matter Institute EMMI, GSI, Planckstr. 1, D-64291 Darmstadt, Germany
	}
	
	\date{\today}
	
	\begin{abstract}
		We study a one-dimensional two-component Fermi gas in a harmonic trapping potential using finite temperature lattice quantum Monte Carlo methods.
        We are able to compute observables in the canonical ensemble via an efficient projective approach.
        Results for density profiles, correlations, as well as energy-related observables are presented for systems with up to $80$ particles and various temperatures.
        Our simulations reproduce known numerical results and compare well against available experimental data close to the ground state, while at higher temperature they are benchmarked against the exact solution of the two particle system.
        This provides an indication that a standard lattice discretization is sufficient to capture the physics of the trapped system.
		In the special case of a spin-imbalanced gas, we find no sign problem in the parameter ranges studied, allowing access without the need of specialized methods.
        This includes simulations close to the ground state and at large population imbalance, where we present results for density correlations, indicating pairing at finite total  momentum.
	\end{abstract}
	
	\maketitle
	
%%%%%%%%%%%%%%%%%%%%%%%%%%%%	
\section{Introduction}
\label{sec:Intro} 

	 Over recent years, considerable advances have been made in the realm of ultracold atoms, both theoretically and experimentally~\cite{PhysRevLett.108.075303,PhysRevA.86.031601,doi:10.1126/science.1240516,PhysRevLett.114.230401}.
	 The unprecedented ability to experimentally tune physical properties of the system, such as the scattering length via a Feshbach resonance~\cite{RevModPhys.82.1225}, the shape of trapping potentials, including  dimensionality~\cite{PhysRevLett.110.200406, Holten:2021pex, PhysRevLett.94.210401}, and the exact number of particles in the trap~\cite{doi:10.1126/science.1201351, bayha2020observing}, has allowed for investigations of a wide range of physical phenomena in strongly correlated systems.
	 In particular, pairing between fermions that experience contact interactions has been a topic of great interest for over a decade~\cite{PhysRevLett.92.160401,PhysRevA.78.033607,PhysRevA.88.063643,PhysRevA.101.033601,PhysRevLett.125.060403}.
	 By tuning the scattering length, the system can be brought from a Bardeen-Cooper-Schrieffer (BCS)  superfluid, where pairs are weakly bound, to strongly bound bosonic dimers, forming a Bose-Einstein-Condensate (BEC), which enables experiments to probe a rich phase structure in the crossover region~\cite{holten2018anomalous, PhysRevLett.114.230401, doi:10.1126/science.1214987}.

     In contrast to BCS-like paring, pairs at finite momentum may form in the presence of  a population imbalance.
	 This has been first studied by Fulde and Ferrell~\cite{PhysRev.135.A550}, and Larkin and Ovchinnikov~\cite{Larkin:1964wok}, and is commonly known as FFLO-type pairing.

	 In this work, we investigate the thermodynamic and pairing properties of a spin $1/2$ fermionic system interacting via a zero range attractive interaction within a harmonic trapping potential in one spatial dimension.
	 We study systems with equal and unequal number of spin-up and spin-down particles, both at finite temperature and in the ground state, which we are able to connect almost continuously.

     The majority of lattice simulations of ultracold quantum gases to date were performed without an external potential and aimed to take a thermodynamic limit.
     They encompass studies of ground state and finite temperature properties in one to three spatial dimensions~\cite{PhysRevA.84.061602, Shi_2015, PhysRevLett.115.115301}, and the thermodynamics of the unitary gas~\cite{PhysRevA.78.023625, PhysRevLett.121.130406} in particular.
     Recently, the BKT-transition temperature~\cite{PhysRevLett.129.076403}, as well as pseudogap effects~\cite{ramachandran2022pseudogap}, were computed in the BEC-BCS crossover regime of the 2D gas.

     Previous theoretical works on the ground state of the trapped system includes exact diagonalization approaches~\cite{PhysRevA.88.033607,PhysRevA.91.043610}, which are usually confined to small particle numbers.
    Non-uniform lattice Monte Carlo methods~\cite{PhysRevA.91.053618, BERGER2016103} have been used to study systems of up to 20 particles,
    while coupled cluster~\cite{PhysRevA.92.061601, Grining_2015} and diffusion Monte Carlo~\cite{PhysRevA.78.033607} approaches have allowed for computations with higher particle numbers.
    At finite temperature, the canonical system has been studied with exact diagonalization~\cite{Pecak:2022dlf}, while lattice-based methods include grand canonical auxiliary field and canonical stochastic Green's function~\cite{PhysRevA.82.013614} approaches.
    Theoretical searches for exotic pairing have been carried out using, amongst others,  Monte Carlo methods~\cite{PhysRevA.78.033607,PhysRevA.82.013614,Rammelmuller:2020vwc,Rammelmuller:2021oxt,Attanasio:2021jkk}, diagrammatic approaches~\cite{Tajima:2020etx}, and exact diagonalization methods~\cite{Pecak:2022dlf}.

	 Lattice simulations often employ periodic boundary conditions, allowing the physical system to enjoy translation invariance.
	 The presence of an external trapping potential breaks this symmetry, making the density profile of the atomic cloud a more interesting object and bringing the theoretical setting closer to its experimental counterpart.	 
     However, simulations at low temperature can suffer from numerical issues due to large separations of scales and require a large temporal extent of the lattice, making them computationally costly.
     The first issue is addressed by employing safe matrix multiplication methods, while the latter is mitigated by the use of a truncation technique that removes unoccupied modes from the simulation~\cite{PhysRevLett.123.136402}.
    Both methods are described in section~\Cref{sec:model}.

%%%%%%%%%%%%%%%%%%%%%%%%%%%%%	
\section{Model and Methods}
\label{sec:model}

The continuum Hamiltonian of the system under investigation is given by 
	\begin{align}
		\hat{H} =&\, \int \text{d}x \, \hat{\psi}^\dagger_\sigma(x) \left( \frac{-\nabla^2}{2m} + \frac{1}{2}m \omega^2 x^2 \right) \hat{\psi}_\sigma(x) \nonumber \\[1ex]
		 & + g \int \text{d}x \,  \hat{n}_\uparrow(x) \hat{n}_\downarrow(x)\,,
	\end{align}
	where $\hat{\psi}_\sigma^\dagger$ and $\hat{\psi}_\sigma$ are, respectively, creation and annihilation operators of fermions in spin states $\sigma \in \{\uparrow, \downarrow\}$, and the particle number density operators are given by $\hat{n}_\sigma = \hat{\psi}^\dagger_\sigma \hat{\psi}_\sigma$.
	In the following we always use $m=1$, leaving the trap frequency $\omega$ and lattice coupling $g_L$ as parameters for the lattice systems. 
	The choice of the lattice coupling $g_L$ has been such that the ground state energy on the lattice agrees with the analytical computation of a $1+1$ particle system with coupling $g$, see \cite{TwoColdAtomsinHarmonicTrap, PhysRevA.91.053618}.
	The given Hamiltonian is put on a rectangular spatial lattice of size $L$ and spacing $a$ with $N_x = L/a$ sites and periodic boundaries.
	The parameter $\omega$ has to be chosen such that the characteristic length scale of the harmonic oscillator,
	\begin{equation}
		L_T = \frac{1}{\sqrt{\omega}}\,,
	\end{equation}
	can be resolved, i.e., $1 \ll L_T/a \ll  N_x$.
	
	The grand canonical partition function of the system is given by 
	\begin{align}\label{eq:partition_function}
		Z &= \mathrm{Tr}\left[ e^{-\beta (\hat{H} - \mu_\uparrow \hat{N}_\uparrow - \mu_\downarrow \hat{N}_\downarrow)} \right]\,,
	\end{align}
	where $\beta = 1/T$ is the inverse temperature, $\mu_\sigma$ is the chemical potential for each spin species, and $\hat{N}_\sigma$ is the respective number operator. To proceed, we cast this expression into a path-integral form by employing a Trotter decomposition,
	discretizing the imaginary time evolution over $N_t$ steps of size $\Delta t = \beta / N_t$.
	Then, by performing a Hubbard-Stratonovich (HS) transformation, we rewrite the partition function of our fermionic system in the form of a path integral over a bosonic field $\phi$,
	\begin{align}\label{eq:path_integral}
		Z &= \int \mathcal{D} \phi \, \det M_\uparrow(\mu_\uparrow, \phi) \det M_\downarrow(\mu_\downarrow, \phi) p(\phi)\,,
	\end{align}
    where $p(\phi)$ accounts for the weight that depends solely on the bosonic field.
	The fermionic matrix is given by $M_\sigma = \mathbb{1} + U_\sigma(\mu_\sigma, \phi)$, where $U_\sigma(\mu_\sigma, \phi)$ are $N_x \times N_x$ matrices containing the information related to the kinetic energy, harmonic trap, and interaction~\cite{PhysRevD.24.2278,PhysRevB.24.4295}.
    In general, we may choose the fields $\phi$ as discrete or continuous, as well as bounded or unbounded.

%%%%%%%%%%%%%%%%%%%%%%%%%%%%%%%%%%%
\subsection{Sampling approach}
	
	We have chosen the HS field to take values $\pm 1$ on every lattice site, leading the path integral in \labelcref{eq:path_integral} to become a sum over a discrete set of configurations, which we sample via the Metropolis algorithm.
	Explicitly, on each spacetime point we employ the well-known density channel transformation~\cite{PhysRevB.28.4059}
	\begin{equation}
		e^{- \Delta t g \hat{n}_{i \uparrow} \hat{n}_{ i \downarrow} } = \frac{1}{2} \sum_{x_i= \pm 1} e^{\left(\gamma x_i - \Delta t g / 2\right) \left(  \hat{n}_{i \uparrow} + \hat{n}_{ i \downarrow} - 1 \right) }\,,
	\label{eq:density_channel}
	\end{equation}
	where the coupling is given by $\cosh(\gamma) = e^{-\Delta t g / 2}$.

    The matrix $U_\sigma(\mu_\sigma, \phi)$ is constructed using Fourier acceleration, i.e., we apply the kinetic energy in momentum space and the harmonic trap and interaction part in position space
    \begin{align}\nonumber 
        B_n(\mu, \phi) = &\,e^{-\Delta t K} e^{-\Delta t V(\phi_n)}\,, \\[1ex]
        U_\sigma(\mu_\sigma, \phi) =&\, \prod_{n=0}^{N_t-1} B_n(\mu_\sigma, \phi)\,.
    \end{align}
    In the definitions above, the chemical potential contribution can be included in either the kinetic or potential part at each step.
    It is not necessary to make the matrix product symmetric here, as the determinant and relevant observables all obey a form of cyclic invariance.
    
    In practice, the costliest part of a simulation lies in constructing the matrices $U_\sigma(\mu_\sigma, \phi)$ and determinants $\det M_\sigma(\mu_\sigma, \phi)$.
    Na\"ively, the computational cost scales as $\mathcal{O}(N_t N_x^2 \log N_x )$ and $\mathcal{O} (N_x^3)$ respectively, which can become prohibitive for large systems or when measuring higher order moments of observables.

    At low temperatures, the matrices $U_\sigma(\mu_\sigma, \phi)$ can become ill-conditioned, particularly when the relative difference between the largest eigenvalues and those around unity (indicating the Fermi surface) cannot be resolved by double precision numbers.
    To address this, we construct the matrices $U_\sigma(\mu_\sigma, \phi)$ using stable matrix multiplication techniques relying on intermediate QR decompositions, which have proven to be both reliable and fast~\cite{PhysRevB.40.506, 10.21468/SciPostPhysCore.2.2.011}. 

    In our simulations, we also exploit the low-rank nature of the determinant as introduced in~\cite{PhysRevLett.123.136402}.
	Unoccupied modes are excluded from the simulation dynamically, resulting in matrices $U_\sigma$ of size $N_x \times n_\sigma$, where $n_\sigma$ is comparable to the number of particles of spin $\sigma$ at low temperatures.
	Typically, $n_\sigma$ is much smaller than $N_x$, leading to considerably faster computations.
    More precisely, the aforementioned cost of $\mathcal{O}(N_t N_x^2 \log N_x )$ is reduced to $\mathcal{O}(N_t N_x n_\sigma \log N_x)$ for the bulk of the computation,
    while the determinant is computed with $\mathcal{O}(n_\sigma^3)$ operations.
    It is worth noting that the first QR decomposition performed when constructing $U_\sigma$ remains an $\mathcal{O}(N_x^3)$ operation, which is gradually reduced to $\mathcal{O}(n_\sigma^2 N_x )$, as modes are removed.
    The truncation procedure is especially convenient in harmonically trapped systems, as the occupation has to be small in comparison to the system size in order to resolve the trapping potential.

    Configurations are proposed via a force bias~\cite{Shi_2015}, resulting in acceptance rates well beyond $90\%$.
    Since a single proposal updates all fields on a single time slice,	configurations are usually decorrelated within only a few sweeps of the whole lattice.

%%%%%%%%%%%%%%%%%%%%%%%%%%%%%%%%
\subsection{Reweighting to the canonical ensemble}
	
All sampling is performed in the grand-canonical ensemble. To obtain observables in the canonical ensemble, we employ a reweighting procedure using Fourier projection~\cite{PhysRevC.49.1422}. The canonical partition function is given by
    \begin{equation}
        Z_N = Tr[\hat{P}_{N_\uparrow} \hat{P}_{N_\downarrow} e^{-\beta  \hat{H}  }] \,,
        \label{eq:canonical_partition_function}
    \end{equation}
	with the projection operator
	\begin{equation}
		\hat{P}_{N_\sigma} = \frac{1}{N_x} \sum_{j=0}^{N_x-1} \exp\left[i \, 2 \pi j \frac{ N_\sigma - \hat{N}_\sigma}{ N_x}\right]\,.
	\end{equation}

    In \labelcref{eq:canonical_partition_function}, we can introduce to the exponential additional terms, $ \beta \mu_\sigma (\hat{N}_\sigma - N_\sigma )$, to stabilise the Fourier sum numerically.
    For practical purposes, the chemical potential is the one we sample the grand canonical system at but has no actual influence on the canonical trace.
    It is tuned such that the average particle number in the grand canonical ensemble is close to the desired one in the canonical ensemble.

	In a general case, the sum over projection angles must be over the full basis set of the system. However, when a truncation is applied, it is enough to consider the truncated basis size~\cite{GILBRETH2021107952}. This leads to a significant decrease in computational cost in dilute systems.
	
    We define a modified matrix for each projection angle in the Fourier sum as $M^{l}_\sigma = \mathbb{1} + \exp \left[- i 2 \pi l  \right] U_\sigma(\mu_\sigma, \phi)$,
    and use $z^{lm}(\mu_\uparrow,\mu_\downarrow,\phi) = \det M^{l}_\uparrow(\mu_\uparrow, \phi) \det M^{m}_\downarrow(\mu_\downarrow, \phi)$ for the weight.
	A measure of the overlap between the canonical and grand canonical partition functions is given by
    \begin{align}
        \langle W_N \rangle 
                            =&\, \left \langle \frac{1}{N_x^2} \sum_{l,m} e^{\frac{i (l N_{\uparrow}+m N_{\downarrow})}{2 \pi N_x}} \frac{z^{lm}(\mu_\uparrow, \mu_\downarrow, \phi)}{z(\mu_\uparrow, \mu_\downarrow, \phi)} \right \rangle \,,\label{eq:overlap}
    \end{align}
    where the expectation value is with respect to the grand canonical weight. We omit the chemical potential prefactor since it only gives a normalization, dropping out of observables.
    Note that the ratio does not necessarily need to be close to unity for good statistics, as the partition functions are not normalized.
    In practice, low ratios do not seem to pose a problem, as long as the relative fluctuations in \labelcref{eq:overlap} are small. 
    This is generally the case when the average particle number of the grand canonical system agrees with the target particle number of the canonical one.
    At higher temperatures, we can usually reweight to a larger range of particle numbers.
    Computation of observables proceeds similarly, now including an additional sum over projection angles
    \begin{align}
		\langle O \rangle_N &= \frac{1}{\langle W_N \rangle N_x^2} \times \nonumber\\
		&\left \langle \sum_{l,m} e^{\frac{i (l N_{\uparrow}+m N_{\downarrow})}{2 \pi N_x}}  \frac{z^{lm}(\mu_\uparrow, \mu_\downarrow, \phi)}{z(\mu_\uparrow, \mu_\downarrow, \phi)} O^{lm}_N(\phi)  \right \rangle \,.
    \end{align}
    For the one body density operator, the observable to be computed is
    \begin{equation}
        n_{ij , N}^{lm}(\phi) = \left[ ( \mathbb{1} + e^{i 2\pi l / N_x} U^{-1} )^{-1}  \right]_{ij} \,,
    \end{equation}
    which is independent of the index $m$, since the different spin species factorize.
    Higher order observables are computed using the one body operator and Wick's theorem in analogy to the grand canonical case.
	To ensure efficient computation of canonical observables the $U$ matrices are diagonalized, again making use of the truncation similar to what is introduced in~\cite{GILBRETH2021107952}.
    This works by taking advantage of the reduced matrix sizes at low particle numbers, extracting only the relevant eigenvalues and eigenvectors.

    In a general situation, for instance at positive coupling $g$ or in the presence of spin imbalance, positivity of the weight is not guaranteed.
    This is accounted for in our simulation by standard reweighting.
    In practice, we always sample the absolute value of the grand canonical weight, and perform the full reweighting in a single step with the projection to the canonical ensemble.
    \begin{figure*}
        \includegraphics[width=\textwidth]{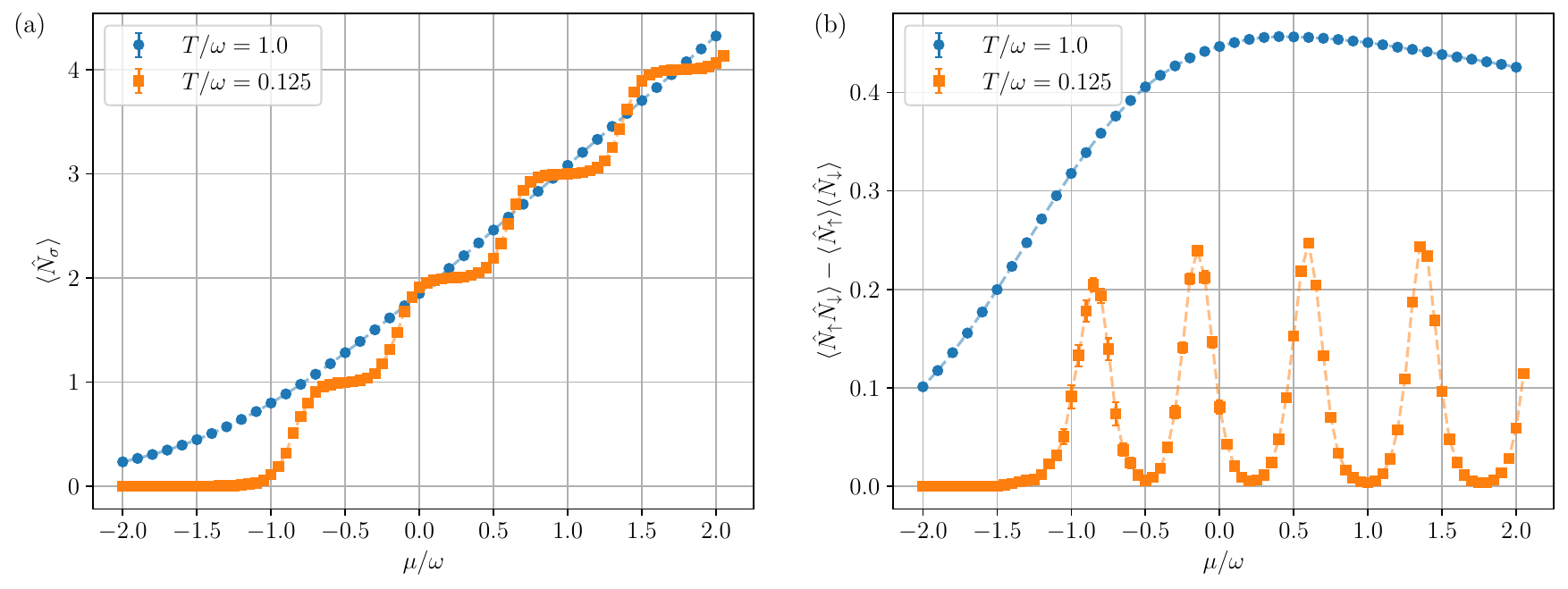}
		\caption{\label{fig:particle_number} (a) Expectation value of the particle number operator as function of the chemical potential for different temperatures for $g/\sqrt{\omega}=-3$.
        (b) Correlation between the particle number operators for each spin species as function of the chemical potential for different temperatures for $g/\sqrt{\omega}=-3$. Dashed lines are drawn to guide the eye.
        }
    \end{figure*}
%

%%%%%%%%%%%%%%%%%%%%%%%%%%%%%%%%%%
	\subsection{Ground state projection}
	
	The finite temperature simulations described above are well suited for the simulations of the system at temperatures close to $T=0$. However, they are not designed for the explicit computation of the ground state, $\Psi_0$.
	To address this, we complement the finite temperature simulations with a ground state projection in the canonical ensemble, following the method described in detail in, e.g.,~\cite{PhysRevB.40.506, ZhangGSP}. 
	This projective method relies on a trial state $\Psi_T$, which is taken to be a Slater determinant, and uses that for non-vanishing overlap $\braket{\Psi_T | \Psi_0} \neq 0$, the trial state converges to the ground state, 
	\begin{align}
		\lim_{\beta \to \infty} e^{-\beta \hat{H}} \ket{\Psi_T} = \ket{\Psi_0}\,.
	\end{align}
	We choose the trial state as the exact solution of the non-interacting system, although more sophisticated options are possible. For the ground state projection, the imaginary time evolution is split into $N_t$ steps of size $\Delta t = \beta / N_t$.
	At each time slice, we apply a symmetric Trotter decomposition and utilize Fourier acceleration when applying them to the states. Again, we make use of QR decompositions to allow for a stable evolution of the Slater determinants.
	Moreover, to handle the Hamiltonian's interaction term, we use the same HS transformation as for the finite temperature simulation from \labelcref{eq:density_channel}. The respective auxiliary fields are again sampled using the force bias method. 
	Altogether, this leads to a ground state projection operator of the form
	\begin{align}
		e^{-\beta \hat{H}} \approx \prod\limits_{n=1}^{N_t} e^{-\Delta t\, K/2} e^{-\Delta\, t V(\phi_n)} e^{-\Delta t\, K/2}\,.
	\end{align}	
	The observables $O$ are then computed using the symmetric estimator
	\begin{align}
		\langle O \rangle_{\mathrm{GSP}} = \frac{\bra{\Psi_T} e^{-\beta \hat{H}/2} \; O \; e^{-\beta \hat{H}/2} \ket{\Psi_T}}{\bra{\Psi_T} e^{-\beta \hat{H}/2}\; e^{-\beta \hat{H}/2} \ket{\Psi_T}}\,.
	\end{align}
This concludes our set-up. 
	
%%%%%%%%%%%%%%%%%%%%%%%%%%%%
\section{Results}

Our results are structured as follows. In \Cref{sec:lattice_parameters}, we provide a brief discussion on the lattice size and parameters.
\Cref{sec:balanced_gas} presents results for density profiles and correlations in systems with balanced population.
In \Cref{sec:energy_observables} we provide a comparison to exact diagonalization results for separation energies and compute the temperature dependence of the pairing gap for various particle numbers.
Finally, in \Cref{sec:imbalanced_gas}, we briefly discuss the sign problem and provide a tomographic picture of density-density correlations in the imbalanced system.

%%%%%%%%%%%%%%%%%%%%%%%
    \subsection{Lattice parameters} \label{sec:lattice_parameters}

    An essential aspect of lattice simulations is the choice of parameters to ensure a rapid convergence towards the continuum limit.
    In our case, there are several limits that need consideration.
    One is the temporal lattice spacing $a_\tau$, which dictates the number of time slices used and thus influences the error of the Trotter decomposition.
    In all situations studied here, we find $a_\tau / a = 0.05$ to be sufficiently small to ensure a negligible Trotter error.

    The ratio $L_t / a$ governs the finite size and finite distance errors and is set to $L_t / a = 4$ for all simulations. Simultaneously, the total number of spatial sites is $N_x=80$, corresponds to $20$ traps lengths.
    We find these parameters to be sufficient for the particle content and couplings studied in the main text. 
    However, it is worth noting that a higher number of particles necessitates a larger spatial extent to avoid finite size effects, while a larger coupling would require a smaller lattice spacing to mitigate finite spacing effects.

    For the system with $1+1$ particles, the continuum Hamiltonian can be diagonalized analytically, yielding an exact comparison for observables in the canonical ensemble.
	To determine the appropriate bare lattice coupling, we tune it such that it yields the exact ground state energy of the two-particle system, given by the relation for the continuum coupling~\cite{TwoColdAtomsinHarmonicTrap,DAmico_2014}
    \begin{equation}
        g(E) = \frac{\sqrt{2} (E-1) \Gamma \left(1-\frac{E}{2}\right)}{\Gamma \left(\frac{3}{2}-\frac{E}{2}\right)}\,.
    \end{equation}
    Achieving this does not require a direct lattice simulation, as the two-particle lattice system can be numerically diagonalized, providing the desired ground state energy.

	\subsection{Spin balanced gas} \label{sec:balanced_gas}
    \begin{figure*}
        \includegraphics[width=\textwidth]{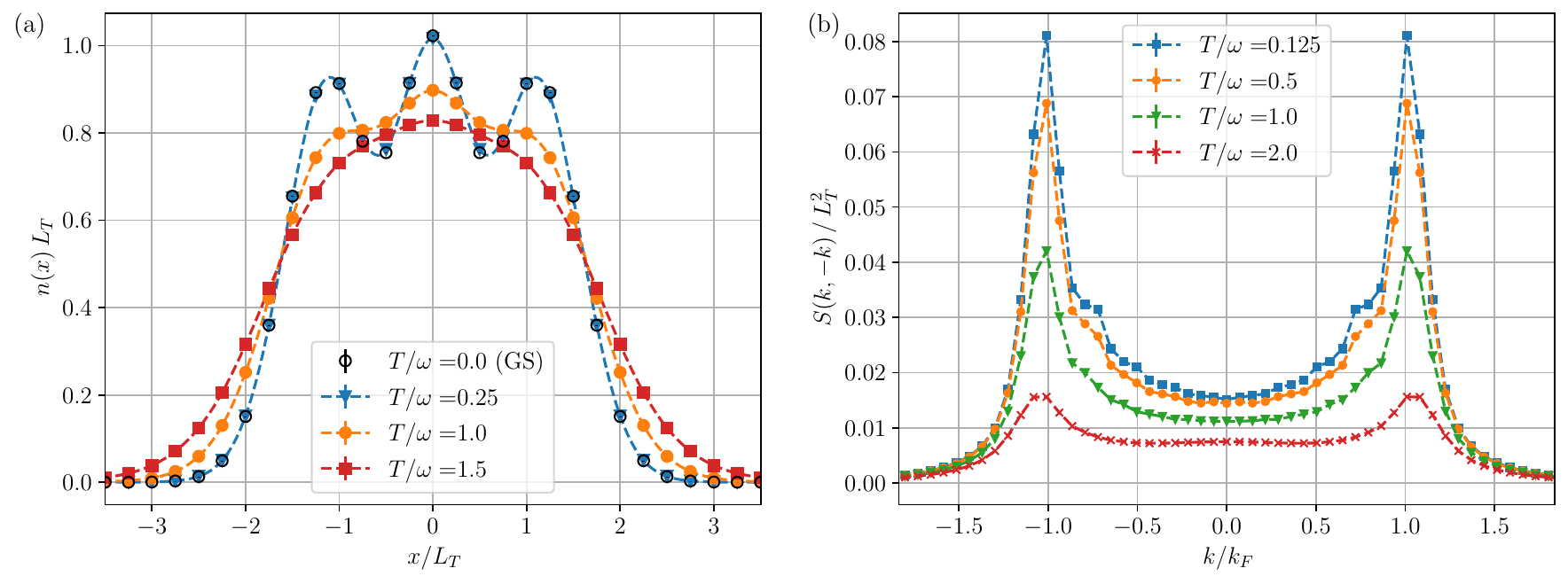}
        \caption{\label{fig:DensityObs} (a) Density profile per spin species for $3+3$ particles and $g / \sqrt{\omega}=-3$ at three different temperatures. Dashed lines are splines through the data points and hollow circles indicate data from projection to the ground state. (b) Anti-diagonal of the density-density correlation function for $10+10$ particles at $g / \sqrt{\omega}=-3$ and four different temperatures. }
    \end{figure*}
	In the case of spin-balanced systems, the determinants in \labelcref{eq:path_integral} are real and equal, such that no sign problem is present.
	This is true for the density channel Hubbard-Stratonovich transformation used in this work, as well as the spin-z channel not applied here, where the determinants are complex conjugate of each other.
		
	Before making further use of the particle number projection methods described above, we computed the expectation value of the particle number operator $\hat{N}_\sigma = \int dx \, \hat{n}_\sigma(x)$ as function of the chemical potential for different temperatures, along with the correlation between up and down spin number operators, $\langle \hat{N}_\uparrow \hat{N}_\downarrow \rangle - \langle \hat{N}_\uparrow \rangle \langle \hat{N}_\downarrow \rangle$.

	\Cref{fig:particle_number} (a) depicts clear steps at nearly integer particle numbers for the lowest temperature of $T/\omega = 0.125$, indicating the thresholds in chemical potential where each energy level is filled.
	This effect is smoothened by thermal fluctuations, as can also be seen.
	Moreover, the attractive interaction between up and down spins causes the departure from the empty system to occur at negative chemical potentials.
	The non-interacting theory, on the other hand, would have this threshold around $\mu/\omega = 1$ for small but finite temperature.

	\Cref{fig:particle_number} (b) shows the (connected) correlation between $\hat{N}_\uparrow$ and $\hat{N}_\downarrow$.
	Similar to what can be seen in the average particle number, this correlation function for $T/\omega=0.125$ exhibits repeating behavior following the filling of the different energy levels.
	Conversely, at high temperature this behavior is not present, and the correlator changes more smoothly with the chemical potential.

    Turning to the canonical ensemble, we investigate the density profile of a fixed number of particles in the trapping potential.
    The ground state shows characteristic particle peaks, originating directly from the wave functions of the harmonic oscillator states.
    These oscillations are already present in the non-interactig system, where the density, in terms of the single particle wave functions $\psi_n(x)$, is given by $n_f(x) = \sum_{n=0}^{N_\sigma - 1} |\psi_n(x)|^2$.
    
    In \Cref{fig:DensityObs} (a), the density profile for $g/\sqrt{\omega}=-3$ with $3+3$ particles is plotted for various temperatures.
    In the ground state, the density distribution displays three peaks, corresponding to the number of particles of each spin species.
    The same behavior is observed at temperatures significantly smaller than the spacing between energy levels in the trapping potential, where the ground state contribution dominates.
    As the temperature increases to the point where the thermal energy becomes comparable to the energy gap, the peaks disappear, and the density profile gradually smoothens.
    
    With larger particle numbers, the particle peaks are expected to decrease in amplitude and become more frequent, eventually converging towards a smooth profile in the thermodynamic limit (see \Cref{sec:largeN}).

   As an indicator for the existence of pairing in the system, we compute the connected density-density correlation function in momentum space, given by
   \begin{equation}
       S(k,k') = \langle n_\uparrow(k) n_\downarrow(k') \rangle - \langle n_\uparrow(k) \rangle \langle n_\downarrow(k') \rangle \,.
       \label{eq:shot_noise}
   \end{equation}
   Pairing around the Fermi surface is expected to manifest as positive correlation peaks at $S(\pm k_F, \mp k_F)$.
   In  \Cref{fig:DensityObs} (b) we present results for the system at $g/\sqrt{\omega}=-3$ and $S(k,-k)$ for various temperatures.
   Close to the ground state the peaks around the Fermi surface are more pronounced.
   The correlations decrease rapidly towards larger momenta, but they do not vanish at vanishing opposite momenta due to the finite particle number and the resulting finite size of the systems, as compared to the thermodynamic limit of an untrapped gas.
   We observe a weakening in correlations, similar to the breakup of ground state features in the density profile, when temperatures become comparable to the level spacing of the trapping potential.
   However, in contrast to the oscillations in the density profile, the density-density correlations at higher temperatures remain clearly visible.
   While the peaks around the Fermi surface are less pronounced, the overall correlation appears flatter at small momenta.

\begin{figure*}[t]
    \includegraphics[width=\textwidth]{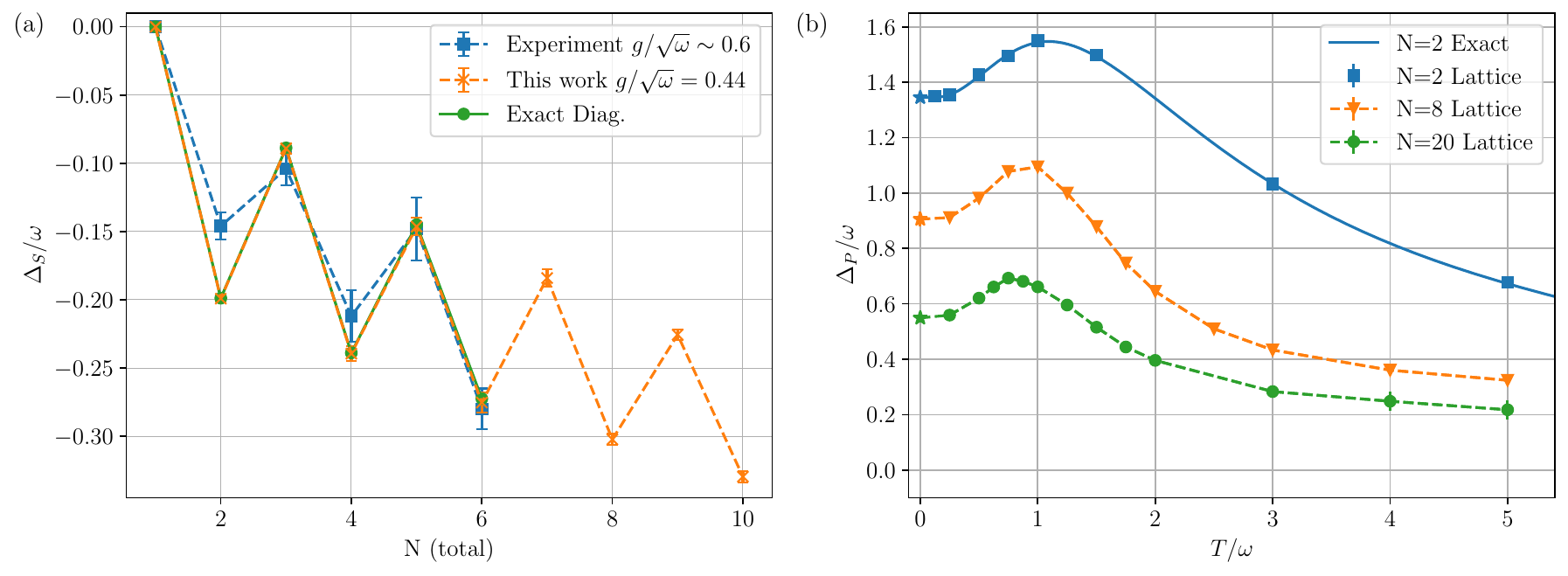}
    \caption{\label{fig:Energy} (a) Separation energies in the ground state for up to $5+5$ particles. Experimental data ~\cite{PhysRevLett.111.175302} (blue squares) and exact diagonalization results~\cite{PhysRevA.91.043610} (green circles) are compared to results from finite temperature lattice simulations at $T / \omega=0.125$ (orange cross). (b) Pairing gap at $g / \sqrt{\omega}=-3$ for $1+1$, $4+4$ and $10+10$ particles at finite temperatures.
    The two particle system is compared to the exact result of the continuum theory (solid line).
    Stars indicate results from ground state projection.}
\end{figure*}

    \subsection{Energy observables} \label{sec:energy_observables}
    Having explored density observables and their correlations, we now shift our focus to energy-related quantities. In particular, separation energies have been measured experimentally~\cite{PhysRevLett.111.175302}
    and studied theoretically via exact diagonalization~\cite{PhysRevA.91.043610, Grining_2015}, and provide a good benchmark for the validity of our computations close to the ground state. 
    The separation energy can be understood as the interaction energy cost of adding a single particle to the system and is defined in terms of the ground state chemical potential of the system as follows:	
	\begin{eqnarray}
		\mu(N) &=& E(N) - E(N-1)\,, \\
		\Delta_S(N) &=& \mu(N) - \mu^*(N)\,,   \label{eq:SepEn}
	\end{eqnarray}
    where $\mu^*(N)$ is the chemical potential of the free system.
		
	At low temperatures, this requires simulations of imbalanced system, since the overlap of the balanced canonical simulation with the imbalanced sector becomes small. 
	In \Cref{fig:Energy} (a) we observe excellent agreement between our results, marked as ``Lattice'' and computed with a finite temperature of $T/\omega=0.125$, and results computed via the exact diagonalization method from~\cite{PhysRevA.91.043610}.
    The step like behavior, with lower energies for even particle numbers, which correspond to shell closures, indicate the presence of pairing between particles of up and down spin.
	We note also the proximity to experimental values at a slightly different coupling.
    Some deviations are visible in comparison to the experiment, likely due to anharmonicity in the trapping potential present in the experiment.
    While beyond the scope of this work, our method generally allows the study of an arbitrarily shaped external potential, which opens up avenues to study effects of anharmonicity in a controlled manner.
    For the comparison to exact diagonalization results we match both two body energies to the same ground state values, resulting in a value for the coupling slightly smaller than what was given in~\cite{PhysRevA.91.043610}.
     While the results in \Cref{fig:Energy} (a) are at small coupling, similar agreement with exact diagonalization is found in more strongly coupled scenarios. 
    
    Next, we compute the energy staggering pairing gap, defined for even particle numbers as:
	\begin{eqnarray}
		\Delta_P (N) &=& \frac{1}{2} [2 E(N/2 - 1, N/2) \nonumber
		- E(N/2, N/2) \\  &&- E(N/2-1, N/2-1)] \,.
	\end{eqnarray}
	This quantity serves as an indicator for pairing in the system, and has been used to study pseudogap effects~\cite{PhysRevLett.125.060403, ramachandran2022pseudogap} and pair correlations~\cite{PhysRevA.88.063643} in higher dimensions.
    Although higher order estimators are available~\cite{PhysRevC.65.014311}, we use a three point estimator here for simplicity.
    At high enough temperatures, it is often enough to sample for a single chemical potential, and reweight the different particle numbers from the same data. 
    This approach breaks down around $T/\omega \sim 0.5$, where we perform independent simulation for the energies at different particle numbers. 

    \begin{figure*}[t]
        \includegraphics[width=\textwidth]{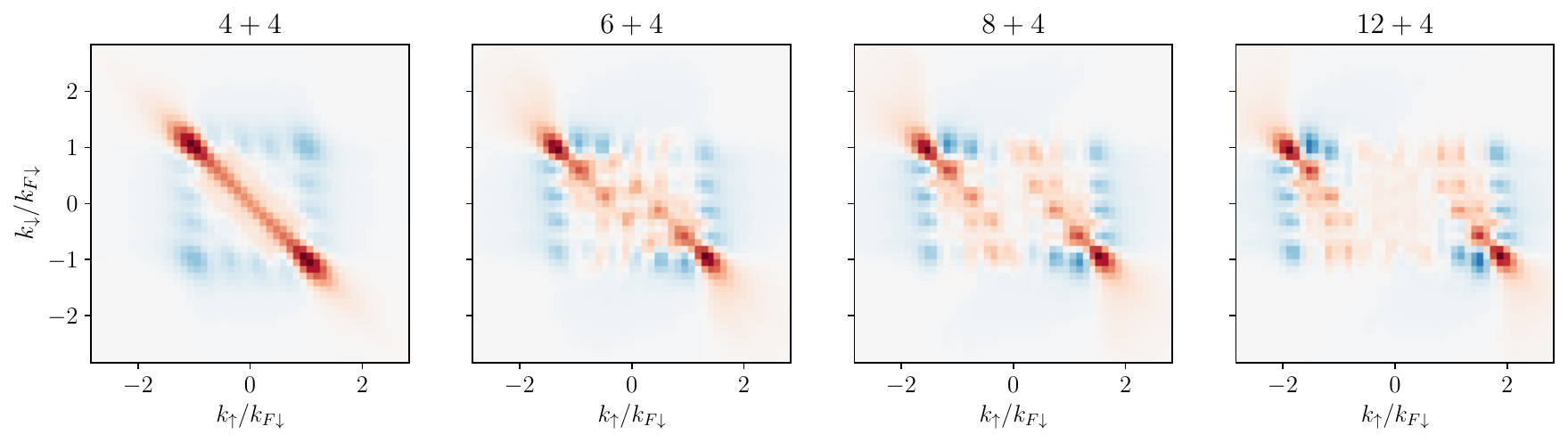}
        \caption{\label{fig:FFLO} Density-density correlations of the system at $g/\sqrt{\omega}=-3$ and $T/\omega=0.25$, with varying imbalance. 
        We observe clear signals of unconventional pairing in the presence of an imbalance. To retain visibility, all color scales are normalized independently to span the full range of values in the corresponding data.}
    \end{figure*}

    \Cref{fig:Energy} (b) depicts the pairing gap at $g / \sqrt{\omega}=-3$ for $1+1$, $4+4$, and $10+10$ particles at finite temperature.
	For two particles we find good agreement with the exact result of the continuum theory up to $T/\omega=5$.
    We see an increase in the gap when increasing temperature, before it decays again. 
	This behavior can likely be explained by the lower degeneracy in the first few excited states of the $1+1$ system compared to its non-interacting counterpart.
    Consequently, the energy of the single particle contribution increases more rapidly than that of the two particle one.
    The good agreement of the two-particle systems with the exact result indicates the validity of the renormalization approach used to fix the coupling, where no excited states were considered. 
    However, we expect the agreement to break down at higher temperatures, as high-energy states beyond the limits of the lattice start to contribute significantly.
    
    The systems with eight and twenty particles respectively follow a similar general trend to the two-particle one, but at lower total magnitude, consistent with available ground state results~\cite{PhysRevA.92.061601}.
    Interestingly, we still find peaks at finite temperature, which move to lower temperature when increasing the particle number.
    It remains unclear, from our computations, whether the peak converges to a fixed temperature or vanishes in the thermodynamic limit, indicating a shell effect.

%%%%%%%%%%%%%%%%%%%%%%%%%%%%    	
\subsection{Imbalanced gas}
\label{sec:imbalanced_gas}
	
    An important open question in the study of ultracold fermionic gases is the existence of an exotic paring phase at finite spin imbalance.
    Such a phase, known as FFLO phase, is characterised by pairing at non-degenerate Fermi surfaces, resulting in pairs with finite total momentum $k_{F \uparrow} - k_{F \downarrow}$.

    It is not clear, however, to what extent such systems can be studied using standard lattice methods.
	In the case of imbalanced spin species, the configuration weight is not necessarily positive anymore, and a sign problem can arise.
	This is the case in both 2D and 3D simulations \cite{PhysRevA.82.053621, PhysRevA.86.023630}.
	In contrast, some previous studies of Fermi gases in with contact interactions in 1D have not shown a sign problem in the considered parameter regions~\cite{Attanasio:2023Qs, PhysRevD.98.054514, PhysRevResearch.3.033180}.
    The trapped system considered here appears to show similar behavior.
    For no temperature, chemical potential or coupling studied do we find negative weight configurations.
	It is important to note that this is not an issue of ergodicity, as our code runs into the expected sign problems in 2D and 3D and reproduces known results in 1D as has already been shown above in the case of separation energies, which require simulations at a slight imbalance.

	We complement previous studies that consider trapped systems with imbalances \cite{Dobrzyniecki:2021fqq, PhysRevA.82.013614, Pecak:2022dlf}, by computing pairing patterns at higher particle numbers than done previously.
    In \Cref{fig:FFLO} we present a visualization of the connected density-density correlations in \labelcref{eq:shot_noise} for a system with $4+4$ to $4+12$ particles at $g/\sqrt{\omega}=-3$ and $T/\omega=0.25$ in momentum space.
    The spin-balanced system is peaked at $k_\uparrow + k_\downarrow=0$, while the imbalanced systems show a clear signal of pairing at finite momentum, which is consistent with the expected FFLO behavior.
    It is important to note that \Cref{fig:FFLO} does not show the relative magnitude of correlations at different particle numbers; each plot is independently normalized to ensure visibility. 
    Moreover, several pockets of positive and negative correlation are visible, forming an oscillatory pattern.
    These oscillations, similar to the oscillatory pattern in the density profile, are a feature of the harmonically trapped system, not found in the same way in the untrapped gas~\cite{Rammelmuller:2020vwc}.  
    While a thorough analysis is left to future work, we generally find a decrease in correlations when going to higher temperatures and larger imbalance,
    which aligns with results from a recent exact diagonalization study of few-particle systems at finite temperature~\cite{Pecak:2022dlf}.

%%%%%%%%%%%%%%%%%%%%%%%%%%%%
\section{Conclusion}

 We have presented results from lattice simulations of trapped fermionic systems in one dimension, in both the case of balanced and imbalanced populations.
    The sampling was performed in the grand canonical ensemble with a reweighting step to give canonical expectation values, which is more efficient than directly sampling the canonical weight but requires tuning the chemical potential.
    The stabilization procedure we used allowed us to simulate the full range of temperatures, down to the ground state, where we compare to a projective approach.
    To verify the validity of our results, we compared to experimental and theoretical data for the separation energies of up to six particles, finding good agreement.
    Additionally, we computed the energy staggering pairing gap, which agrees with exact results for the two particle system and is also computed for up to twenty particles. 
    The computation of separation energies in particular requires simulations for spin imbalanced systems, which we find to be sign problem free in the parameter ranges studied.
    This is also the case when computing density-density correlations in the presence of larger imbalances, where we find clear signals of unconventional pairing.

    In future studies, it may be interesting to explore polaronic effects in the systems, as no sign problem appears to be present, allowing computations even at large imbalances.
    In this paper we put a focus on attractive contact interactions, due to the absence of a sign problem.
    However, recent work on the application of complex Langevin methods in the study of both ultra-cold fermions~\cite{Rammelmuller:2020vwc, Attanasio:2021jkk, Rammelmuller:2021oxt, PhysRevA.98.053615, PhysRevD.95.094502} and bosons~\cite{PhysRevA.101.033617, PhysRevA.106.063308, heinen2023simulating}, may allow for the study of repulsive interactions in the future.
    While computationally more costly, the lattice approach generalises straightforwardly to higher dimensions, making studies of trapped gases in 2D possible in the future.\\[0ex]
    %	
%

%%%%%%%%%%%%%%%%%%%%%%%%%%%%%%%%%%%%%%%%%%%%%%%
\begin{acknowledgments}

We thank P.~D'Amico for discussions and data. We also thank S. Shokri and M. Haverkort for discussions. This work is funded by the Deutsche Forschungsgemeinschaft (DFG, German Research Foundation) under Germany’s Excellence Strategy EXC 2181/1 - 390900948 (the Heidelberg STRUCTURES Excellence Cluster) and the Collaborative Research Centre SFB 1225 - 273811115 (ISOQUANT). We also acknowledge support by the state of Baden-W\"urttemberg through bwHPC.

\end{acknowledgments}

%%%%%%%%%%%%%%%%%%%%%%%%%%%%%%%%%%%%%%%%%%%%%%%
\appendix
\section{Lattice dispersion}
\begin{figure}[t]
    \includegraphics[width=0.48\textwidth]{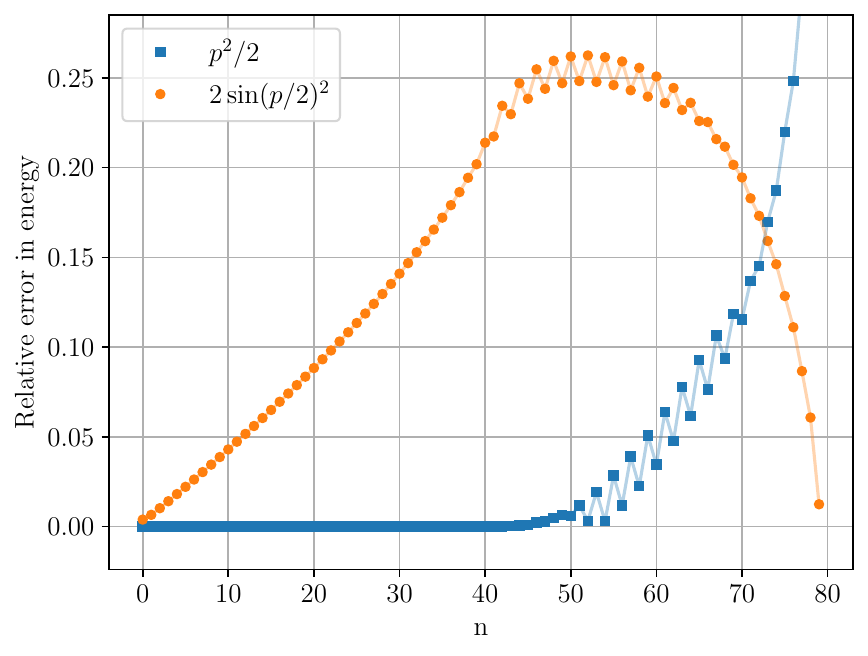}
    \caption{\label{fig:Dispersion} Relative errors of the energy levels with $p^2$ dispersion and Hubbard dispersion in the non-interacting system.}
\end{figure}

The discretization of the Laplace operator is an important aspect of lattice simulations. In particular, a well-chosen dispersion can reduce finite distance effects and improve convergence to the continuum limit.
In this work, we use
\begin{equation}
    E(p) = \frac{p^2}{2} \,,
\end{equation}
for the kinetic energy of the lattice system.
Another form often used is given by the finite difference approximation of the derivatives, yielding
\begin{equation}
    E(p) = 2 \sin^2 \left( \frac{p}{2} \right) \,,
\end{equation}
which is the standard dispersion of the Hubbard model.
As a simple check, we diagonalize the non-interacting lattice Hamiltonians with both dispersions and compare the resulting energies to the continuum theory.
The lattice parameters from the main text are used, which are $N_x = 80$, $\omega = 0.0625$.
In \Cref{fig:Dispersion}, we plot the relative errors in the energy levels of the harmonic trap.
The quadratic dispersion shows significantly better agreement, with no significant deviations up to half filling, while the Hubbard dispersion deviates even for the lowest lying states.
In both cases, we observe an even-odd effect in higher shells. This is not an issue for our simulations, as they are carried out in more dilute regimes.
It should be noted that methods which keep the energy levels exact are available, but require a non-uniform spatial lattice~\cite{PhysRevA.91.053618,BERGER2016103}.

\section{Large particle numbers} \label{sec:largeN}
\begin{figure*}
    \includegraphics[width=\textwidth]{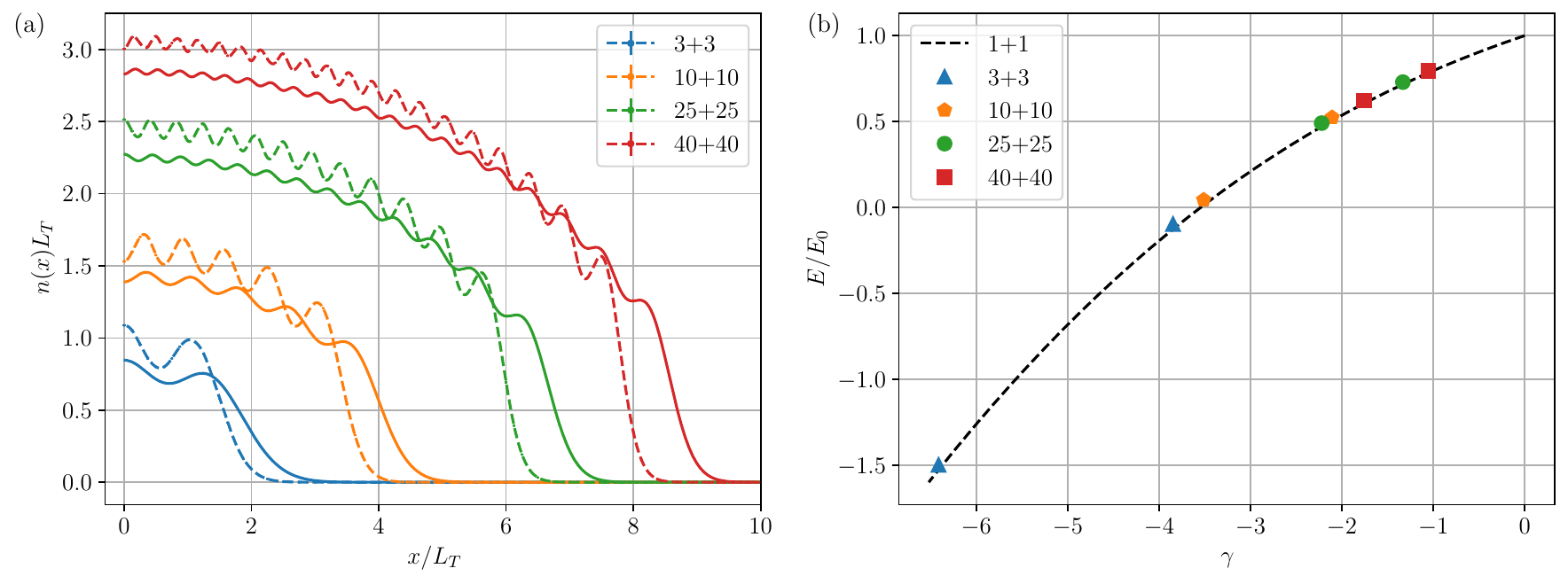}
    \caption{\label{fig:LargeN} (a) Density profiles per spin species for $N=6$, $20$, $50$ and $80$ particles at $g / \sqrt{\omega} = -5$ and $T/\omega=0.25$. Splines connecting the data points (dashed lines) are drawn to guide the eye.
    The ground state densities of the corresponding non-interacting systems are plotted as solid lines.
    (b) Energy of the system, normalized by the non-interacting energy, as function of the rescaled coupling $\gamma$.
    The energies of the two-particle system in the ground state are plotted as a dashed line.}
\end{figure*}

To show the applicability of our approach to systems of larger particle number, we compute the density profiles and energies for systems of up to $N=80$ particles.
To this end, we adjust the lattice parameters to be $N_x=200$ and $\omega = 0.015625$, which corresponds to $L_T = 8$.
In \Cref{fig:LargeN} (a) we plot the density profiles $N=6$, $20$, $50$ and $80$ particles at $g / \sqrt{\omega} = -5$ and $T/\omega = 0.25$, as well as the ground state densities of the corresponding non-interacting systems.
The larger number of lattice sites, and smaller lattice spacing, allow for a satisfactory resolution of the density oscillations close to the ground state.
To further verify the computations, we compare the energies obtained at different particle numbers.
In \cite{PhysRevA.92.061601}, the authors found the energy, normalized by the non-interacting energy, to only show very weak dependence on the particle number when plotted against the rescaled coupling
\begin{equation}
    \gamma = \frac{\pi g}{\sqrt{\omega N}} \,.
\end{equation}
This behavior is reproduced by our data, as can be seen in \Cref{fig:LargeN} (b), where we compare the energies against those of the two-particle system in the ground state.
The lattice results generally overshoot the two-particle system slightly, which is expected at higher particle number, but may also partially result from the finite temperature.

Our current approach is somewhat limited in the couplings that can be studied, as stronger interactions lead to ergodicity issues in the Markov process.
This makes tuning the particle number challenging, in particular at very low chemical potentials.
Similar issues are encountered in the Hubbard model at half filling, where the issue can be mitigated by global updates~\cite{PhysRevB.44.10502}.
Alternatively, it may be possible to circumvent the problem by sampling the canonical weight more directly.
 
\bibliography{refs}
		
\end{document}